\documentclass[twocolumn,hyperpdf,amsmath,amssymb,aps,prd,10pt,superscriptaddress,nofootinbib,noeprint,preprintnumbers,floatfix]{revtex4-2}

\usepackage{graphicx, color}
\usepackage[dvipsnames]{xcolor}
\usepackage{booktabs}
\usepackage{graphicx}

\usepackage[letterspace=-10]{microtype} 

\usepackage{bm, amsmath, amsfonts, amssymb,xfrac}
\usepackage{multirow, tabularx, dcolumn, soul}
\usepackage{mathtools, leftidx, braket, slashed, cancel, bigdelim}
\usepackage{blkarray}
\usepackage[figures]{rotating}

\usepackage{tikz}
\usetikzlibrary[plotmarks]
\usepackage{anyfontsize}

\usepackage[utf8]{inputenc} 
\usepackage{hyperref}
\definecolor{jlab_red}{RGB}{192,39,45}
\definecolor{jlab_orange}{RGB}{249,102,0}
\definecolor{jlab_blue}{RGB}{47,122,121}
\definecolor{jlab_green}{RGB}{65,125,10}
\definecolor{jlab_gray}{gray}{0.6}
\definecolor{magenta}{rgb}{0.5, 0, 0.5}

\newcommand\bef{\begin{figure}}
\newcommand\eef[1]{\label{fg:#1}\end{figure}}
\newcommand\beq{\begin{equation}}
\newcommand\eeq[1]{\label{#1}\end{equation}}
\newcommand\beqa{\begin{eqnarray}}
\newcommand\eeqa[1]{\label{#1}\end{eqnarray}}
\newcommand\bet{\begin{table}}
\newcommand\eet[1]{\label{tb:#1}\end{table}}
\newcommand\fgn[1]{Figure~\ref{fg:#1}} 
\newcommand\eqn[1]{Eq.~(\ref{#1})}
\newcommand\tbn[1]{Table~\ref{tb:#1}}

\hypersetup{%
pdftitle = {title},
pdfsubject = {},
pdfkeywords = {},
colorlinks = {true},
filecolor = {black},
linkcolor = {jlab_blue},
menucolor = {black},
citecolor = {jlab_green},
urlcolor = {jlab_green},
}{}

\begin{document}

%
\title{Study of isoscalar scalar $bc\bar u\bar d$ tetraquark $T_{bc}$ from lattice QCD}
\author{Archana Radhakrishnan}
\email{archana.radhakrishnan@tifr.res.in}
\affiliation{Department of Theoretical Physics, Tata Institute of Fundamental Research, \\ Homi Bhabha Road, Mumbai 400005, India }
\author{M. Padmanath}
\email{padmanath@imsc.res.in}
\affiliation{The Institute of Mathematical Sciences, a CI of Homi Bhabha National Institute, Chennai, 600113, India}
\author{Nilmani Mathur}
\email{nilmani@theory.tifr.res.in}
\affiliation{Department of Theoretical Physics, Tata Institute of Fundamental Research, \\ Homi Bhabha Road, Mumbai 400005, India }

\preprint{IMSc/2024/2 TIFR/TH/24-2}

\date{\today}
\begin{abstract}
We present a lattice QCD study of the elastic $S$-wave $D\bar{B}$ scattering
in search of tetraquark candidates with explicitly exotic flavor
content $bc\bar u\bar d$ in the isospin $I\!=\!0$ and $J^P=0^+$
channel. We use four lattice QCD ensembles with dynamical $u/d$, $s$,
and $c$ quark fields generated by the MILC Collaboration. A non-relativistic 
QCD Hamiltonian, including improvement coefficients up to $\mathcal{O}(\alpha_sv^4)$, 
is utilized for the bottom quarks.
For the rest of the valence quarks we employ a relativistic overlap action. Five different valence 
quark masses are utilized to study the light quark 
mass dependence of the $D\bar{B}$ scattering amplitude. 
The finite volume energy spectra are 
extracted following a variational approach. The elastic $D\bar{B}$ scattering 
amplitudes are extracted employing L\"{u}scher's prescription. 
The light quark mass dependence of the continuum 
extrapolated amplitudes suggests an attractive interaction between the $\bar B$ 
and $D$ mesons.  
At the physical pseudoscalar meson mass ($M_{ps}=M_{\pi}$) the $D\bar{B}$  scattering amplitude has a 
sub-threshold pole corresponding to a binding energy of $-39(^{+4}_{-6})(^{~+8}_{-18}) \mbox{~MeV}$ with
respect to the $D\bar{B}$ threshold. 
The critical $M_{ps}$ at which the $D\bar{B}$
scattering length diverges and the system becomes unbound corresponds
to $M^*_{ps}=2.94(15)(5) \mbox{~GeV}$. This result can hold significant experimental relevance in 
the search for a bound scalar $T_{bc}$ tetraquark, which could well be the next 
``doubly heavy'' bound tetraquark to be discovered with only weak decay modes.
\end{abstract}

\maketitle

\section{Introduction}
  \label{Intro}
The study of exotic hadrons is one of the most prominent areas of research in contemporary particle physics.
The proliferating list of discovered exotic hadrons, exhibiting properties that demand interpretations 
beyond conventional meson and/or baryon models, continues to captivate the scientific interest. A compilation 
of various exotic hadrons discovered till now and their properties can be found in Ref. \cite{QWGEH}. Among 
these exotic hadrons, those with manifestly exotic flavor content having four quarks in their valence 
structures are particularly popular. Several of them are proximal to open flavor thresholds pointing to a
connection with the corresponding scattering channel for their existence, and possibly their nature. These 
four quark states could be compact tetraquarks, or meson-meson molecular excitations, or a mixture of
both or something more intriguing: a much enviable research topic at this time. An in-depth understanding 
of the binding mechanism governing these hadrons can play a crucial role in elucidating the non-perturbative 
QCD dynamics.

A particularly notable common feature among the discovered four quark hadrons is the presence of at least one 
heavy quark constituent in their valence structure. Phenomenologically it has been hypothesized and discussed 
that a color-singlet combination of two very heavy quarks (anti-quarks) and two light anti-quarks (quarks) can 
form a $QQ\bar{q}_1\bar{q}_2$ bound state  \cite{PhysRevD.15.267,PhysRevD.17.1444}. Recently a handful of 
calculations using first principles method of lattice QCD also strongly indicate the presence of deeply bound 
states with the quark contents $bb\bar{q}_1\bar{q}_2$;\, $q_1\in u, d$; $q_2\in d(s), u(s)$ \cite{Bicudo:2015kna,
Francis:2016hui,Bicudo:2017szl,Junnarkar:2018twb,Leskovec:2019ioa,Hudspith:2023loy}. Very interestingly a 
doubly-charmed four quark hadron, coined as $T_{cc}^{++}$, with the quark content $cc\bar{u}_1\bar{d}_2$ and 
unusually long lifetime, has recently been discovered by LHCb ~\cite{LHCb:2021vvq}. Lattice QCD calculations have 
also investigated $T_{cc}^{++}$ and suggested that the existence of this hadron could be the result of a delicate 
fine tuning between the light and heavy quark masses \cite{Collins:2024sfi,Lyu:2023xro,Chen:2022vpo,Padmanath:2022cvl}.
In summary, lattice QCD calculations and phenomenological investigations consistently suggest the existence of 
deeply bound states in doubly bottom four quark system, referred to as $T_{bb} \in bb\bar{q}_1\bar{q}_2$, 
while experimental evidence has been reported for a four quark hadron ($T_{cc}$) with the quark content 
$cc\bar{u}\bar{d}$. Notably, the charm quark mass is comparatively lighter than the bottom quark mass, 
suggesting potentially differing binding strengths for doubly-bottomed and doubly-charmed four quark states 
due to QCD dynamics operating at multiple scales.

In this respect, four quark systems ($T_{bc} \in bc\bar{q}_1\bar{q}_2$) which are in between $T_{bb}$ and $T_{cc}$, 
{\it i.e}., with a bottom and a charm valence quarks, are of particularly interesting. Phenomenologically, 
predictions on the existence of such states are ambiguous with their energies exhibiting considerable spread 
over several hundreds of MeV around the relevant two-meson threshold. Several model studies based on heavy quark 
symmetry~\cite{Eichten:2017ffp,Braaten:2020nwp,Ebert:2007rn} suggest no binding. Numerous non-chiral models
\cite{Lipkin:1986dw,Zouzou:1986qh,Silvestre-Brac:1993zem,Semay:1994ht,Lee:2009rt,Karliner:2017qjm,PhysRevD.99.014006,
Park:2018wjk,Lu:2020rog,Guo:2021yws,Richard:2022fdc,Song:2023izj} also suggest either a weak binding or an unbound 
system. However, certain chiral models~\cite{Sakai:2017avl,Deng:2018kly,PhysRevD.101.014001,Tan:2020ldi} and QCD 
sum rule investigations~\cite{Chen:2013aba,PhysRevD.99.033002,PhysRevD.101.094026,Agaev:2020zag,Wang:2020jgb} 
propose a more pronounced binding.

In such a scenario, first principles lattice QCD calculations can provide much needed information on the bindings of $T_{bc}$ states.
However, previous lattice QCD calculations~\cite{Francis:2018jyb,Hudspith:2020tdf,Meinel:2022lzo} claim either no evidence for a bound 
$T_{bc}$ state or insufficient statistics to conclude its existence. In Ref.~\cite{Meinel:2022lzo}, the authors do not 
come up to a conclusion due to large uncertainties. In a recent work, we investigated the $J^P=1^+$ channel considering 
chiral as well as continuum extrapolations and found an attractive interaction between the $\bar{B}^*$ and $D$ mesons 
indicating the possible existence of a bound $T_{bc}$ state with a binding energy of $-43(^{+6}_{-7})(^{+14}_{-24})$ 
MeV with respect to the $D\bar{B}^*$ threshold \cite{Padmanath:2023rdu}. Afterwards, in another recent calculation
\cite{Alexandrou:2023cqg}, some of the authors of Ref.~\cite{Meinel:2022lzo} studied $J^P=0^+$ as well as $1^+$ channels, 
involving bilocal two-meson interpolators corresponding to the elastic excitations of $D\bar B^{(*)}$ scattering. They 
found attractive interactions in both channels, and subsequently pointed to the existence of shallow 
bound states. Given the coarse lattices they utilize for these hadrons with two heavy quarks, it will be important to 
check whether the binding of these states observed in Ref.~\cite{Alexandrou:2023cqg}  will survive or enhance with continuum extrapolation.

Motivated by the recent progress, and building upon our previous work for $J^P=1^+$ channel \cite{Padmanath:2023rdu}, 
in this work we perform a lattice QCD calculation of elastic $D\bar{B}$ mesons \footnote{We assume $m_u=m_d$, ignore 
QED effects, and refer to the degenerate ($D^+B^-,~D^0\bar{B}^0$) threshold as $D\bar B$.} scattering in the isoscalar 
scalar channel $I(J^P)=0(0^+)$. Following a partially quenched approach, we investigate the light quark mass dependence 
of the $D\bar B$ mesons scattering. The lattice-extracted scattering amplitudes, employing L\"uscher's finite-volume 
prescription, are extrapolated to the continuum limit. The amplitude at the physical pion mass is deduced following 
a study of the light quark mass dependence of these continuum-extrapolated results. Finally, the hadronic pole 
information in this physical amplitude is studied towards identification of bound state poles.

Experimentally, $J^P=0^+$ channel is also more interesting as it could be the next ``doubly heavy" tetraquark to 
discover since it has a reduced heavy diquark mass that is lower than that for the $bb\bar{q}_1\bar{q}_2$ system. 
It would also likely be the first tetraquark that would unambiguously decay only weakly.

The remainder of the manuscript is structured as follows. A brief overview of our lattice setup is 
provided in Section \ref{sec:lattice}.  In Section \ref{sec:2ptIO}, we discuss various relevant
technical details involved in our calculation such as the observable measured, the interpolating 
operators utilized and the extraction of finite volume energy spectra, which are presented in 
Section \ref{fvresults}. The extraction of scattering amplitudes, continuum extrapolations, and 
chiral extrapolations made are presented in Section \ref{Ampfits}. 
In section \ref{comparison} we present a discussion on the bindings 
of $T_{bc}$ four-quark states, in perspectives of available lattice and non-lattice results, along with 
a comparison of scattering lengths for $DD^*$, $D\bar{B^*}$,  $BB^*$ and $D\bar{B}$ scatterings. 
Finally we summarized our results in section \ref{discuss}.

\section{Lattice Details}\label{sec:lattice}

The computational setup used in this calculation is similar to the one in several of our 
previous publications \cite{Junnarkar:2019equ,Junnarkar:2018twb,Basak:2014kma,
Padmanath:2017lng,Basak:2012py,Basak:2013oya,Mathur:2016hsm,Mathur:2018epb,Mathur:2018rwu,
Junnarkar:2022yak,Mathur:2022ovu} and most recently in Ref.~\cite{Padmanath:2023rdu}.
We use four $N_f=2+1+1$ ensembles with dynamical quark fields respecting a Highly Improved 
Staggered Quark (HISQ) action generated by the MILC collaboration \cite{MILC:2012znn}. 
Other relevant details of various lattice QCD ensembles used are listed in \tbn{lat}. 
The lattices have different volumes and lattice spacings $a$, which are estimated using 
the $r_1$ parameter \cite{MILC:2012znn}. The gauge fields respect one-loop and follow 
the tadpole-improved Symanzik gauge action with tuned coefficients through 
$\mathcal{O}(\alpha_sa^2, n_f\alpha_sa^2)$ \cite{Follana:2006rc}. 
 
The valence quark masses upto the charm quark are realized using an overlap fermion action 
that is $\mathcal{O}(am)$ improved  \cite{Chen:2003im,xQCD:2010pnl}. The bare charm quark 
mass on each ensemble was tuned using the kinetic mass of spin averaged $1S$ charmonia 
$\{a\overline M_{kin}^{\bar cc} = 0.75 aM_{kin}(J/\psi) + 0.25 aM_{kin}(\eta_c)\}$ 
determined for the respective ensembles following the Fermilab prescription \cite{El-Khadra:1996wdx} 
(for more details see  Refs. \cite{Basak:2012py,Basak:2013oya}). The bare strange quark 
mass is tuned  to the physical point such that the lattice estimate for the fictitious 
pseudoscalar $\bar ss$ equals 688.5 MeV \cite{Chakraborty:2014aca}. 

Our setup assumes an exact isospin symmetry $m_u=m_d$ over a range of light quark masses 
corresponding to $M_{ps}\sim$0.5, 0.6, 0.7 (equivalent to the strange quark mass), 1.0, 
and 3.0 (equivalent to the charm quark mass) GeV, to map the light quark mass dependence 
over a wide range of $M_{ps}$ values. In \fgn{ampi}, we present the landscape of different 
light quark masses (in terms of $M_{ps}$) studied in the lattice ensembles employed. We 
utilize a wall-smearing procedure at the quark source for our propagator measurements 
which is described in Refs. \cite{Mathur:2018epb,Junnarkar:2018twb,Mathur:2022ovu}.

We use a nonrelativistic QCD (NRQCD) Hamiltonian approach for the bottom quark~\cite{PhysRevD.46.4052}. 
The bottom quark mass was tuned following the Fermilab prescription~\cite{El-Khadra:1996wdx}, 
matching the lattice-determined kinetic mass of the spin-averaged $1S$ 
bottomonium state to its experimental value.
For details 
regarding the NRQCD Hamiltonian, improvement coefficients, and 
bottom quark mass tuning specific to our setup, see Ref.~\cite{Mathur:2016hsm}.

\bet[tbh]
  \begin{center}
      \begin{tabular}{p{2cm}p{1.6cm}p{2.0cm}p{1.8cm}p{0.6cm}}
      \hline
      Ensemble & Symbol & Lattice Spacing & Dimensions  & $M_{ps}^{sea}$ \\
       & ($a$) [fm] & ($N_s^3\times N_t$)  & \\ \hline

      $S_1$ & \pmb{\textcolor{red}{\tikz{\pgfsetplotmarksize{0.8ex}\pgfuseplotmark{diamond}}}} & 0.1207(11) & $24^3\times64$  & 305 \\
      $S_2$ & \pmb{\textcolor{magenta}{\tikz{\pgfsetplotmarksize{0.8ex}\pgfuseplotmark{pentagon}}}} & 0.0888(8) & $32^3\times96$  & 312\\
      $S_3$ & \pmb{\textcolor{blue}{\tikz{\pgfsetplotmarksize{0.7ex}\pgfuseplotmark{o}}}} & 0.0582(4) & $48^3\times144$  & 319\\
      $L_1$ & \pmb{\textcolor{OliveGreen}{\pgfsetplotmarksize{0.7ex}\tikz{\pgfuseplotmark{square}}}} & 0.1189(9) & $40^3\times64$ & 217 \\   \hline
      \end{tabular}
  \end{center}
  \caption{Details of lattice QCD ensembles employed. $M_{ps}^{sea}$ refers to the sea pion mass. 
      $S_1$, $S_1$, and $S_1$ refer to small spatial volume ensembles and $L_1$ refers to the large volume ensemble.}
\eet{lat}

\bef[tbh]
\includegraphics[width=0.4\textwidth]{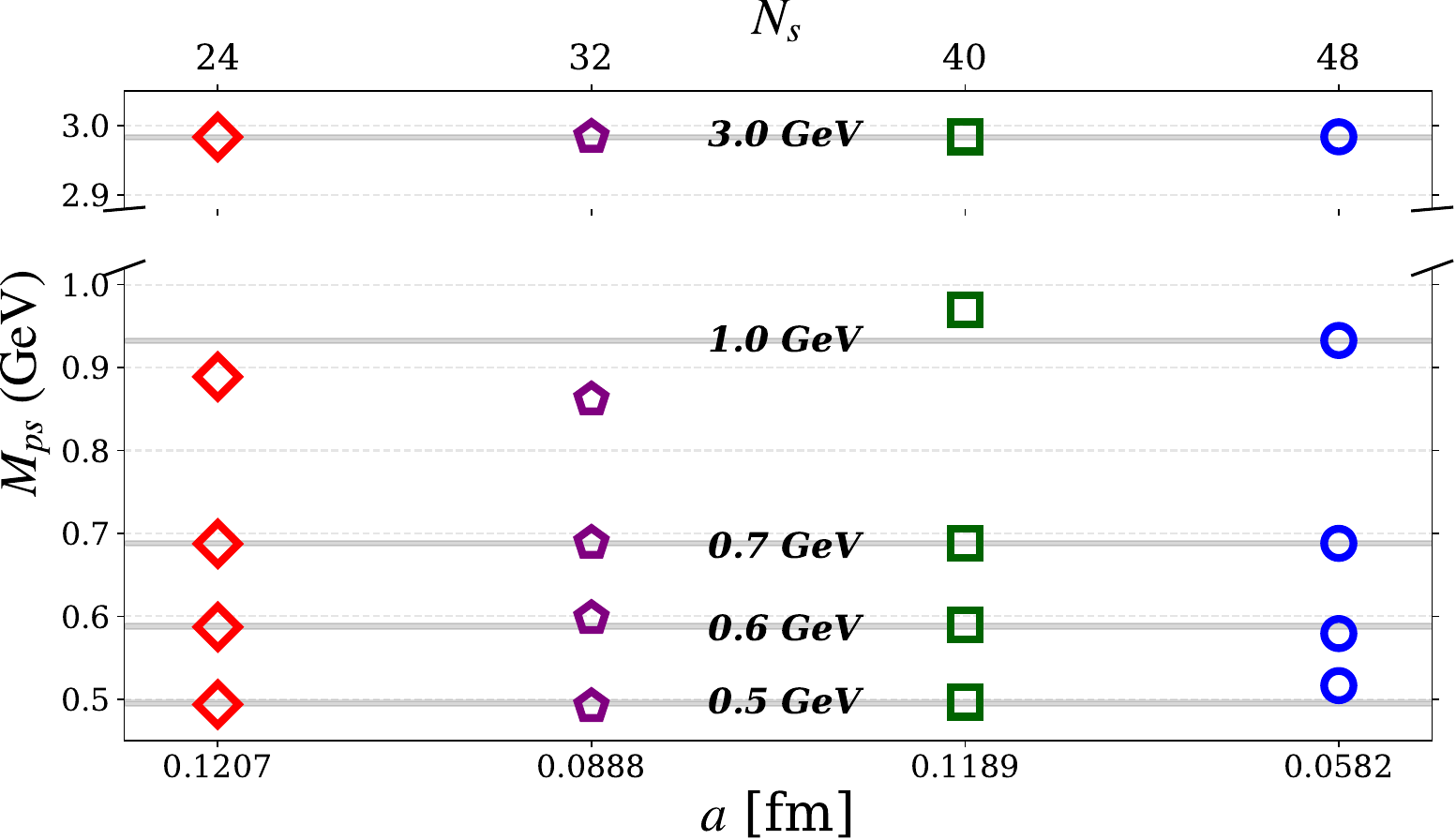}
\caption{A landscape plot of the pseudoscalar masses used across the different lattice ensembles. 
The light quark masses are varied across these five values, while charm and bottom quarks are 
tuned to their physical values. The horizontal gray bands represent estimates of $M_{ps}$, 
to facilitate a comparison of pseudoscalar meson masses across all four ensembles.}
\eef{ampi}

\section{Determining the Finite Volume Spectra using Lattice QCD}\label{sec:2ptIO}
The time dependence of Euclidean two-point correlation functions 
\beq
\mathcal{C}_{ij}(t) = \sum_{\mathbf{x}}\left<\mathcal{O}_i(\mathbf{x},t)\mathcal{\tilde O}_j^{\dagger}(0)\right> = \sum_n \frac{Z_i^n\mathcal{Z}_j^{n\dagger}}{2E^n} e^{-E^nt}.
\eeq{twoptc}
 featuring operators $\mathcal{O}_i(\mathbf{x},t)$ with the desired quantum numbers dictate the time evolution of finite volume spectral levels. 
Here the operator-state overlap $Z_i^n = \bra{0}\mathcal{O}_i\ket{n}$ determines the coupling of the operator $\mathcal{O}_i$ with the state $n$.
The wall-smearing at the quark source in our setup filters out all the high-momentum modes at the source, whereas at the sink time slice we 
utilize a point sink for the quark fields and project the correlation function to its rest frame as shown in \eqn{twoptc}. This asymmetric 
nature of the wall-source point-sink setup is emphasized in \eqn{twoptc} with different operators and the operator-state overlaps at 
the source and at the sink. 

For the $\bar{B}$ and $D$ mesons, we compute two-point correlation functions using the standard local quark 
bilinear interpolators ($\overline Q~\Gamma~q$) with spin structure $\Gamma\sim\gamma_5$. Since we are 
only interested in the rest frame ground state, single meson correlation functions are evaluated only for 
the $A_1^-$ irrep in the finite volume.  

Our study focuses on the $S-$wave $D\bar{B}$ scattering in the rest frame leading to infinite volume quantum numbers 
$J^P=0^+$, which reduces to the $A_1^+$ finite-volume irrep. The elastic two-meson threshold is at 
$E_{D\bar{B}}=m_D + m_{\bar B}$, whereas the lowest inelastic threshold corresponds to the $D^*\bar{B}^*$ 
scattering channel, which is sufficiently high to assume a purely elastic $D\bar{B}$ scattering in $S-$wave.
There are no relevant low lying three particle thresholds in this channel and the lowest multi-particle 
inelastic threshold corresponds to $D\bar B\pi\pi$. 

In the present analysis, we use both a meson-meson type of operator and a local diquark-antidiquark kind 
of operator as in Ref.~\cite{Hudspith:2020tdf}.
\beqa
\mathcal{O}_1(x) &=& [\bar u(x) \gamma_5 b(x)][\bar d(x) \gamma_5 c(x)]  \nonumber \\&& - [\bar d(x) \gamma_5 b(x)][\bar u(x) \gamma_5 c(x)] \nonumber \\
\mathcal{O}_2(x) &=& (\bar u(x)^T \Gamma_5 \bar d(x) - \nonumber \\&& \bar d(x)^T \Gamma_5 \bar u(x)) ( b(x) \Gamma_5 c(x)).
\eeqa{allops}
Here $\mathcal{O}_1(x)$ is a meson-meson operator associated with the $D\bar{B}$ threshold with the individual $D$ 
and $\bar{B}$ forming a color singlet. We do not include any other scattering operators since the next one, 
corresponding to the $D^*\bar{B}^*$ is sufficiently higher up in energy and is assumed to have negligible effects 
on the low-lying spectrum. Excited elastic two-meson operators of $D\bar{B}$ system with nonzero relative meson 
momenta, such as those used in Ref.~\cite{Alexandrou:2023cqg}, are also not utilized in this study. The wall-smearing 
setup we utilize disallows construction of such elastic scattering operators. 

$\mathcal{O}_2(x)$ is a local diquark-antidiquark type operator where all the (anti)quark fields are 
jointly projected to zero momentum. In the color space, diquarks/antidiquarks are built in the 
antitriplet/triplet representations of $SU(3)_c$. In \eqn{allops}, $\Gamma_k = C\gamma_k$ with 
$C=i\gamma_y\gamma_t$ being the charge conjugation matrix. Phenomenologically, doubly heavy tetraquarks 
are expected to be deeply bound and compact in the heavy quark limit, which motivates the use of 
this operator \cite{Francis:2016hui,Czarnecki:2017vco}. Such compact local operators were also considered 
in our previous study \cite{Padmanath:2023rdu} of axial-vector bottom-charm tetraquarks. It is 
also empirically known from several other studies of doubly-bottom tetraquarks that such operators 
have a rich overlap with the ground state 
\cite{Hudspith:2020tdf,Meinel:2022lzo,PhysRevD.107.114510,PhysRevD.100.014503,PhysRevD.99.034507,PhysRevD.96.054510,PhysRevLett.118.142001,PhysRevD.93.034501}.

\bef[h]
\includegraphics[scale=0.47]{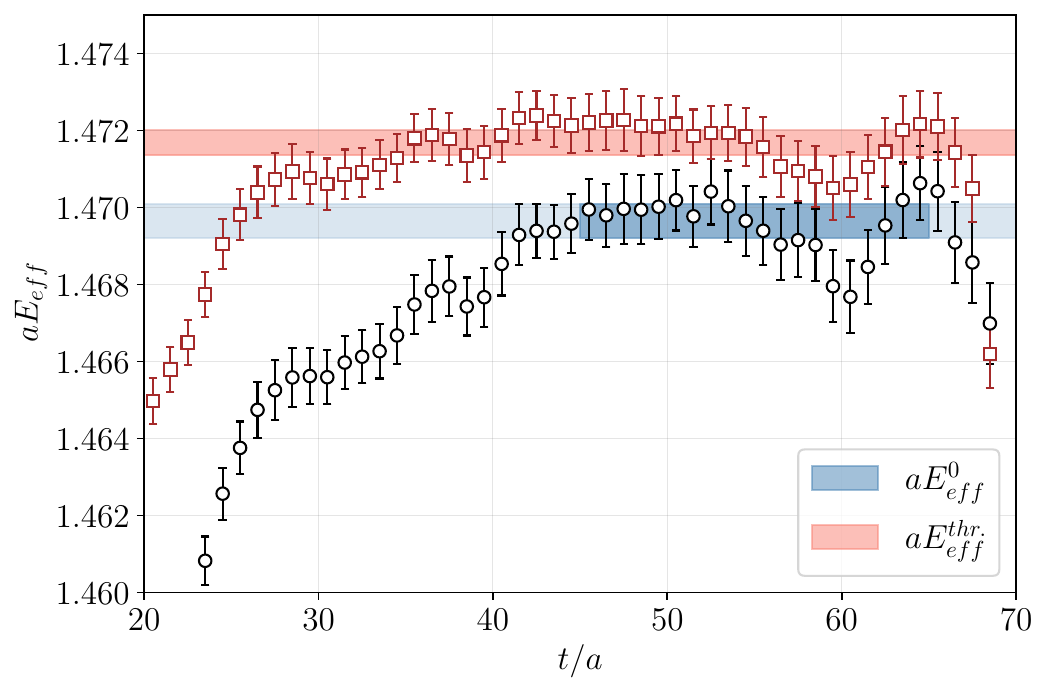}
\caption{Effective energy plot for the eigenvalue of the ground state $\lambda^0(t)$ in circular black 
markers and for the lowest threshold which is the product of single-meson correlators for the $D$ and 
$\bar{B}$ meson, $\mathcal{C}_{D}(t)\mathcal{C}_{\bar B}(t)$ in square red markers. The corresponding blue and 
orange bands are the energy fit estimates using single exponential fit forms on $\lambda^0(t)$ and 
the single-meson correlation functions respectively.}
\eef{effmass}
With these two operators, we find a suitable linear combination that overlaps maximally to the ground 
state by solving for the Generalized Eigenvalue Problem (GEVP) \cite{Michael:1985ne},
\beq
\mathcal{C}(t)v^n(t) = \lambda^n(t) \mathcal{C}(t_0)v^n(t),
\eeq{gvp}
The eigenvalues, $\lambda^n(t)$ correspond to the $n^{th}$ lowest eigenstates with energy $E^n$, where 
$n\le 1$ in our case. We are only interested in the ground state $E^0$. The time evolution of the lowest 
eigenvalue, $\lim_{t\to\infty}\lambda^0(t) \sim A_0e^{-E^0t}$, gives the value of the ground state 
energy in the large time limit, whereas the magnitude of the operator-state-overlaps 
\beq
Z_i^{0}=\bra{0}\mathcal{O}_i \ket{0} = \sqrt{2E^0}(V^{-1})_i^0 e^{E^{0}(t_0)/2},
\eeq{ovp}
indicates the coupling of the operators to the ground state. Here $V$ is the matrix of eigensolutions 
$v^n(t)$, which are expected to be time independent in the large time limit, where $\mathcal{C}(t)$ is saturated 
by the lowest $N$ eigenstates. 

The quality of signals in the energy estimates are assessed using the effective energies, 
\beq
aE_{eff} = [ln(\mathcal{C}(t)/\mathcal{C}(t+\delta t))]/\delta t,
\eeq{meffdef} 
\textit{c.f.} \fgn{effmass}, where we plot present $aE_{eff}$ for the case $M_{ps}\sim3$ GeV on the finest lattice. The black points represent
the effective energy of the interacting system $\mathcal{C}(t)=\lambda^0(t)$, whereas the red points indicate the effective energy of the correlator 
$\mathcal{C}(t) = \mathcal{C}_D(t)\mathcal{C}_{\bar B}(t)$ of the noninteracting system of $D$ and $\bar{B}$ mesons and serve as a reference. A negative shift of the interacting 
energy level with respect to the noninteracting ones is evident in \fgn{effmass}.

The energy estimates are extracted from the correlator data from fitting them with their 
expected asymptotic forms. This can be performed in two ways: the obvious way of fitting 
the interacting correlator $\lambda^0(t)$ directly, or to fit the ratio of correlators
\beq
R^0(t)=\frac{\lambda^0(t)}{\mathcal{C}_{D}(t) \mathcal{C}_{\bar B}(t)},
\eeq{rr1}
with a single exponential form in the large time limit. We are primarily interested in determining 
the energy splittings between the interacting data and the noninteracting one, 
$\Delta E^0 = E^0-M_{D}-M_{\bar B}$. Fits to $R^0(t)$ directly leads to the estimates for $\Delta E^0$.
Alternatively, these splittings can be evaluated as differences between the estimates for energy $E^0$
from fits to $\lambda^0(t)$ and for ($M_D$, $M_{\bar B}$) from separate fits to $\mathcal{C}_{D}(t)$ 
and $\mathcal{C}_{\bar B}(t)$, respectively. A comparison of estimates from these two procedures 
assures that the ground state energy splittings we extract are not influenced by any conspired 
cancellation of noises leading to any fake energy plateaus. We present a demonstration of such
a comparison in \fgn{compr}, where it is evident that the value of $\Delta E^0$ estimated 
from the two different procedures agree with each other within error-bars. This trend is 
observed throughout all the correlators examined. The final results quoted in this paper 
are based on fitting the ratio correlators defined in \eqn{rr1}.

\bef[h]
\includegraphics[scale=0.47]{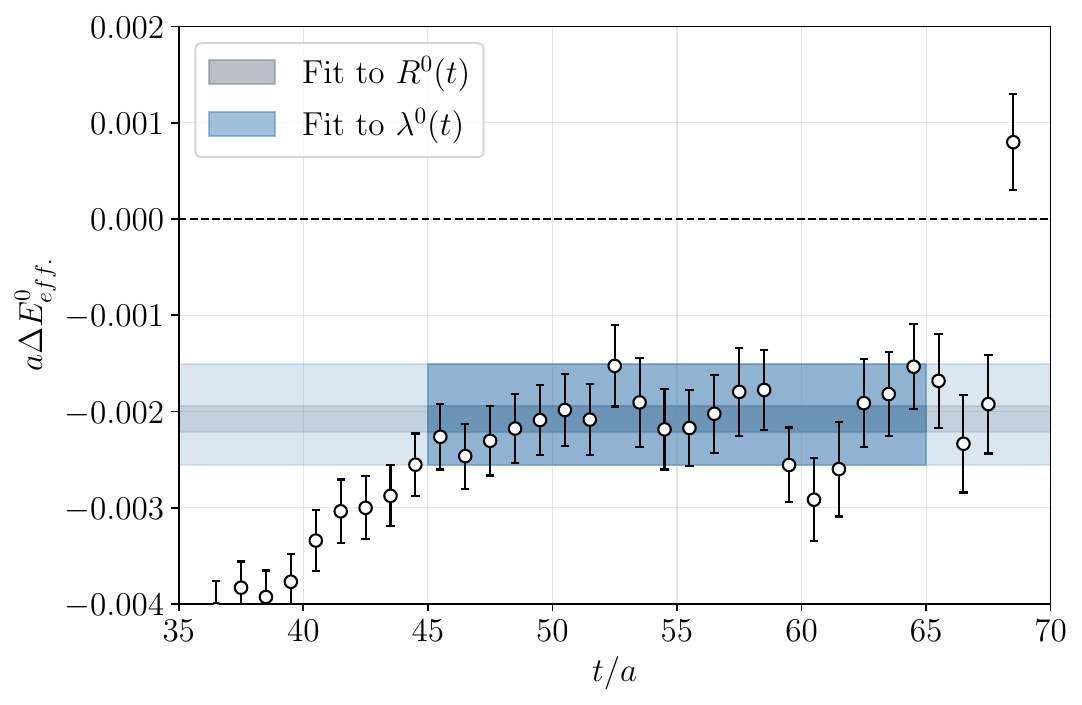}
\caption{$a \Delta E^0$ vs. $t/a$ plot, for $M_{ps}\sim3$ GeV on the finest lattice. Here $a \Delta E^0(t)$, shown in the circular black data data-points, 
is the effective energy splitting determined using \eqn{meffdef} with $\mathcal{C}(t)=R^0(t)$. 
The fit estimates determined from the single exponential fits to $\lambda^0(t)$ and $R^0(t)$ is shown in 
grey and blue bands respectively.}
\eef{compr}

\section{Finite Volume Spectra}\label{fvresults}


\begin{figure*}[thb!]
\includegraphics[scale=0.4]{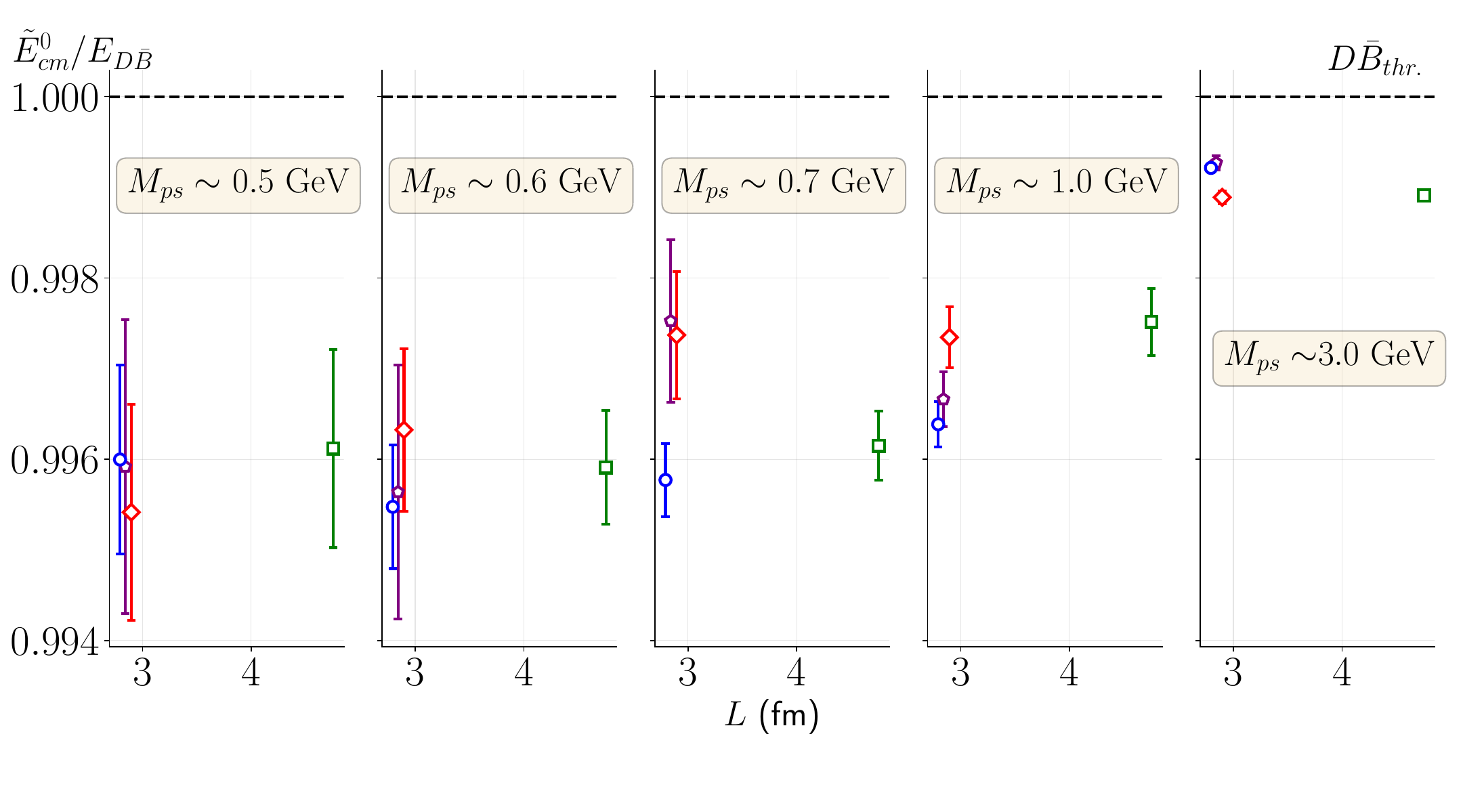}
\caption{The ground state finite volume energies in the $0(0^+)$ $bc\bar u\bar d$ channel. Different panels stand 
for different $M_{ps}$ values indicated on the top of the respective channel. The $y$-axis indicates the energy 
in the center-of-mass frame, in units of energy of the $D\bar{B}$ threshold. The $x$-axis in each panel indicates 
the spatial extent of the lattice ensembles used.  }
\label{fg:spec}
\end{figure*}

In \fgn{spec}, we present the extracted finite-volume energy spectra of the $0(0^+)$ $bc{\bar{u}}{\bar{d}}$ 
channel on the four ensembles listed in \tbn{lat}, at the five different $m_{u/d}$ values corresponding to 
roughly, $M_{ps}\sim$ 0.5, 0.6, 0.7, 1.0, and 3.0 GeV. The energy spectrum shown is normalized by 
the threshold $M_D+M_{\bar B}$, such that center-of-mass energy at threshold is unity in these units. In each 
panel, the $x$-axis represents the spatial extension of the lattice. 

The finite-volume energies are determined from energy splittings extracted from the ratio correlators 
given in \eqn{rr1}. These energy splittings are free of the additive offsets, inherent to the NRQCD formulation, 
as the numerator and denominator in \eqn{rr1} carries same number of valence NRQCD-based bottom quarks. 
The reconstruction of the finite-volume energies from the energy splittings follow the same lines as 
in Ref.~\cite{Padmanath:2023rdu}.

A clear trend of negative shifts for the ground state energies with respect to the $D\bar{B}$ threshold can 
be observed for all the lattices and for all the quark masses studied. It is also evident that this negative 
shifts decreases in magnitude with increasing $M_{ps}$, as expected for a doubly heavy tetraquark system 
\cite{Francis:2016hui,Czarnecki:2017vco}. The variation in this splitting across different lattice
spacings for any given $M_{ps}$ is not transparent due to large uncertainties, whereas unlike in our 
study of axial-vector $T_{bc}$ tetraquark, a moderate trend of decreasing splitting with increasing volume 
can be observed as expected. However, it is too early to substantiate this behavior considering 
the large uncertainties. Despite the large uncertainties, the consistent negative shifts clearly point to 
an attractive interaction between the $D$ and $\bar{B}$-meson in the scalar channel.

In the wall-smearing setup we use, the elastic $D\bar{B}$ excitations involving nonzero relative meson 
momenta are suppressed. This should not affect the ground state determination because it is unlikely that 
operators with relative momenta contribute to the ground state. Additionally, we employ various cross checks 
that helps us estimate the excited state contaminations in the ground state energy, that are then included 
in the systematic uncertainties. We refrain from using or plotting the excited states determined from 
the solutions of \eqn{gvp} in \fgn{spec}, as they do not represent the the elastic $D\bar{B}$ excitations 
in the wall-smearing setup. Another significant limitation of the wall-smearing setup is its asymmetry, 
leading to the possibility of the ground-state energy plateau being approached from below. The agreement 
observed between the energy splittings calculated from ratios of correlators and those determined from 
the difference in energy fit estimates for individual single meson and interacting two-meson correlators 
indicates that our correlator-based fitting estimates effectively manage contaminations from excited states 
which are then incorporated into the systematic errors.

\section{$\mathbf{D\bar{B}}$ scattering amplitude from the finite-volume spectra}\label{Ampfits}
In this section, we present $S$-wave elastic $D\bar{B}$ scattering amplitudes determined following L\"uscher's 
finite-volume prescription~\cite{Luscher:1990ux}. We use only the ground state energies to constrain 
the amplitudes, since the wall-smearing procedure that we utilize for quark sources is not suited to 
extract the elastic excitations, but only the ground states \cite{Padmanath:2023rdu}. For the scalar 
channel considered in this work, the lowest inelastic threshold is $D^*\bar{B}^*$, which is significantly 
high in energy and there are no higher partial wave that can mix with the $S$-wave, justifying an elastic $S$-wave analysis.

A topical aspect in the study of doubly heavy hadrons is the influence of left-hand cuts (lhc) due to 
off-shell pion exchanges\cite{Du:2023hlu}. Recently, there has been efforts to accommodate the lhc effects 
arising from single pion exchanges \cite{Raposo:2023oru,Meng:2023bmz,Hansen:2024ffk}. In $D\bar{B}$ scattering, 
the closest non-analyticity below the threshold can happen from an off-shell two-pion exchange, which has 
its branch point well below the elastic threshold. Hence we ignore any effects of such left-hand nonanalyticities in our analysis. 

\subsection{Amplitude fits using L\"uscher's finite-volume formalism}

The  L\"uscher's finite-volume formalism relates the amplitude of two-particle scattering to 
the finite volume-spectrum in a cubic box. Particularly for the elastic $S$-wave scattering of $B$ and $D$ mesons,
\beq
k\cot[\delta_0(k^2)] = \frac{2Z_{00}[1;(\frac{kL}{2\pi})^2)]}{L\sqrt{\pi}},
\eeq{pcotd}
where $Z_{00}$ is the generalized zeta function described in Ref. \cite{Luscher:1990ux}, $L$ is 
the spatial extent of the cubic box and $\delta_0(k)$ is the $S$-wave phase shift as a function of 
$k$, which is the momentum of either mesons in the center of momentum frame related to the center of 
momentum energy $E_{cm}=\sqrt{s}$ through $4sk^2 = (s-(M_{D}+M_{\bar B})^2)(s-(M_{D}-M_{\bar B})^2)$. From 
\eqn{pcotd}, it is clear that there is a one-to-one correspondence between the energy level and 
the $\delta_0(k)$, \emph{i.e.} each finite-volume energy level provides a specific value of 
the $S$-wave elastic phase shift with which one can constrain the energy or $k$ dependence of 
the phase shift.  

\begin{figure}[h]
  \includegraphics[scale=0.56]{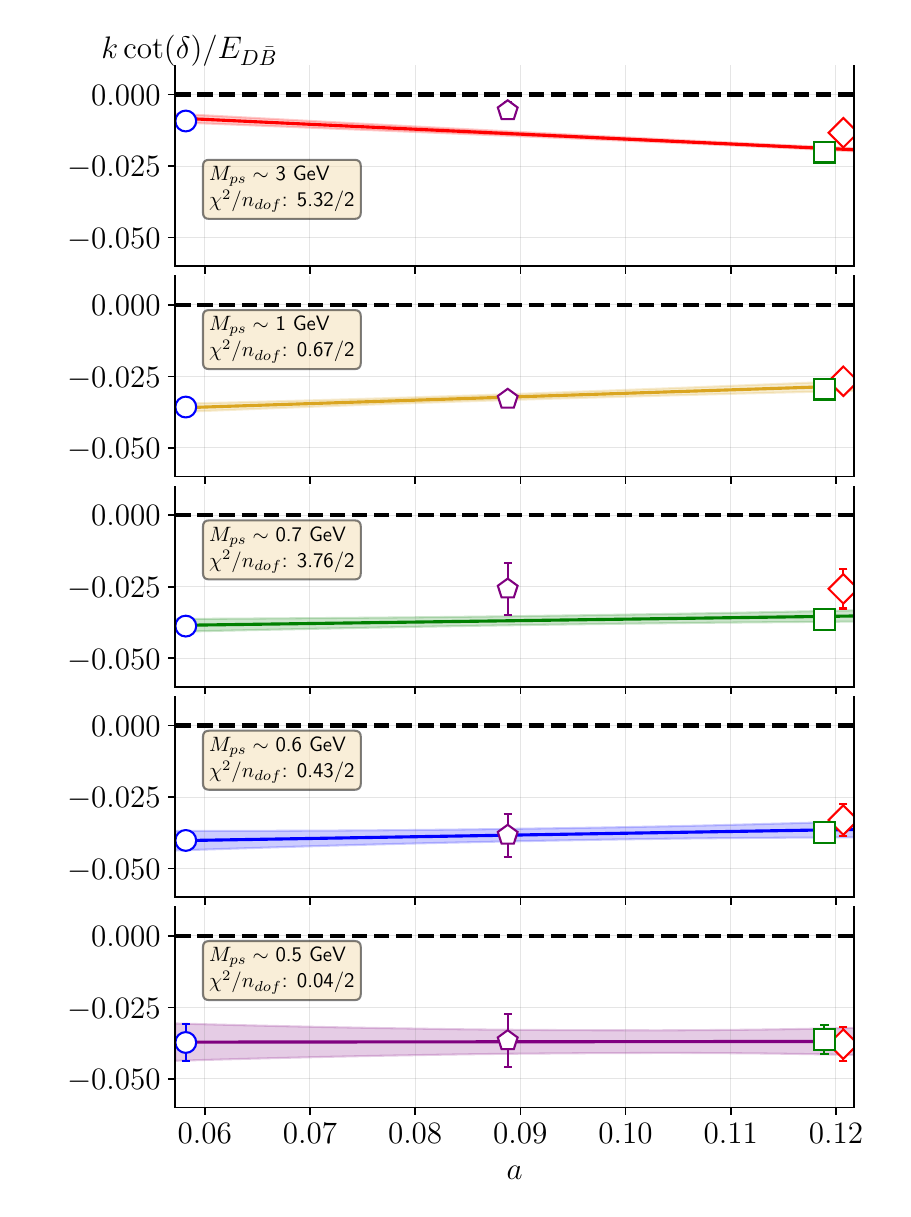}
  \caption{$k\cot[\delta_0]$ normalized by the elastic threshold $E_{D\bar{B}}$, versus lattice spacing, $a$, 
  for the $M_{ps}$ values studied in this analysis as indicated in the different panels. The colored 
  bands indicate the fit results to the amplitude parameterization given in \eqn{prmztn}. The marker 
  conventions are as listed in~\tbn{lat}. For all the $M_{ps}$ values except when $M_{ps}\sim$ 3.0 GeV, 
  the fits show a positive scattering length $a_0$, indicating an attractive interaction.} 
  \eef{kcotd}

We perform the amplitude fits with the ground states from all four ensembles listed in~\tbn{lat}, 
and repeat this for all five values of $M_{ps}$ indicated in \fgn{ampi}. The fits follow minimization of a 
cost function defined as  
\beqa
\chi^2 &=& \sum_{i,j}(f(k_i^2)-f(\{A\},k_i^2)) \nonumber \\
    &~&~~~~~ (\mathcal{C}^{-1})_{ij}(f(k_j^2)-f(\{A\},k_j^2))
\eeqa{chisq}
where $f(k_i^2)$ is the amplitude (lhs of \eqn{pcotd}) extracted from the simulations at $k_i^2$, and 
$f(\{A\},k_i^2)$ is the parameterization of the energy dependence of the amplitude. $\mathcal{C}$ is 
the covariance matrix defined as in Ref.~\cite{Prelovsek:2020eiw}. We verify that the results determined 
from the $\chi^2$ defined in \eqn{chisq} are consistent with that one gets from the procedure outlined 
in Appendix B of Ref.~\cite{Padmanath:2022cvl}. Considering the smallness of $(k/E_{D\bar{B}})^2$ and 
the limited energy range over which the ground states are placed, we assume a zero-range approach for 
the amplitude parameterization. Additionally, we include a linear lattice spacing dependence to account 
for the cutoff effects in the extracted amplitude, which takes the form 
\beq
k\cot[\delta_0] = A^{[0]} + aA^{[1]}
\eeq{prmztn}
where $A^{[0]}=-1/a_0$, with $a_0$ being the scattering length in the continuum limit.

\bet[hb]
  \centering
  \begin{tabular}{clll}
    \toprule
    $M_{ps}$ [GeV] & $\chi^2/d.o.f$ & $A^{[0]}/E_{D\bar{B}}$ & $A^{[1]}/E_{D\bar{B}}$ \\
    \midrule
    0.5 & 0.04/2 & $-0.038(_{-11}^{+15})$ & $~0.004(_{-134}^{+122})$  \\
    \midrule
    0.6 & 0.43/2 & $-0.044(_{-7}^{+8})$  & $~0.06(7)$ \\
    \midrule
    0.7 & 3.76/2 & $-0.042(_{-4}^{+5})$ & $~0.05(_{-4}^{+5})$  \\
    \midrule
    1.0 & 0.67/2 & $-0.043(4)$  & $~0.12(4)$  \\
    \midrule
    3.0 & 5.32/2 & $~0.002(3)$  & $-0.17(_{-3}^{+2})$  \\
    \bottomrule
  \end{tabular}
  \caption{Fit results for amplitude with parameterization given in \eqn{prmztn} at various 
  light quark masses, corresponding to $M_{ps}$ in the first column. The optimized parameter 
  values in the table are expressed in units of the energy of the threshold, $E_{D\bar{B}}$. 
  The numbers within the parenthesis indicate the statistical errors.}
\eet{Ampfits1}

In Figures \ref{fg:kcotd} and \ref{fg:pcotdelta_boot}, we present the fit results to $k\cot[\delta_0]$ 
(the bands) as a function of the lattice spacing and $(k/E_{D\bar{B}})^2$, respectively, along with the lattice 
data. The bands in \fgn{pcotdelta_boot} are the continuum extrapolated results given by the parameter 
$A^{[0]}$. Different horizontal panels represent different $M_{ps}$ values. The best fit parameters and 
corresponding quality of fits are tabulated in~\tbn{Ampfits1}.

Given the negative energy shifts and the sign of $A^{[0]}=-1/a_0$, 
determines the nature of the near-threshold poles, if any. Note that for the non-charm 
light quark masses, $a_0$ is consistently positive suggesting that the strength of interaction to be 
sufficient enough to house a bound state. Whereas at the charm point $a_0$ is negative, despite negative 
energy shifts, suggesting only a feeble interaction that cannot hold a subthreshold pole with square-integrable 
wave-function. This is similar to our observation in the axial-vector channel using the same setup and 
formalism in Ref.~\cite{Padmanath:2023rdu}, as well as to the phenomenological expectation for doubly 
heavy four quark systems, here the binding energy is expected to decrease with increasing light quark 
masses for fixed heavy quark masses.

Another interesting observation is on the variation in the cut off dependence of the amplitudes as 
the light quark masses are varied. The cut off dependence is accounted by the parameter $A^{[1]}$, 
which shows a signature change as the light quark mass increases towards the charm point. This 
suggests that for a doubly heavy four quark ($QQ'l_1l_2$) system with 
$(m_{l_1} = m_{l_2}, ~m_{Q},m_{Q'}>>m_{l})$, the cut off effects weaken the finite-volume energy
splitting of the ground state with the elastic threshold. On the other hand, close to the charm 
point (where $m_{Q},m_{Q'}\sim m_{l}$) such effects enhance this energy splitting in the $QQ'l_1l_2$ 
system determined in a finite-volume. Relatively large errors at the non-charm $M_{ps}$ values 
partially obscure these effects, if any exist, while at the charm point such effects are clearly 
reflected. Any further quantified comments on this lattice spacing dependence is currently beyond 
the scope of the current work, particularly considering the large uncertainties.

\subsection{Extrapolation to physical light quark mass}
Following the extraction of the continuum extrapolated amplitude at different $M_{ps}$ values, we delve 
into the light quark mass dependence of the fitted parameters. The leading order $M_{ps}$ term in 
the chiral expansion suggests the $M_{ps}$ dependence of hadron masses for light $m_{u/d}$ values 
($m_q\lesssim\Lambda_{QCD}$) to be $M_{ps}^2$. Whereas in the heavy light quark mass regime 
($m_q>>\Lambda_{QCD}$) heavy hadron masses are expected to be linear in $M_{ps}$~\cite{Neubert:1993mb}. 
With these phenomenological expectations, we use three fit forms like~\cite{Padmanath:2023rdu},
\beqa
	f_l(M_{ps}) &=& \alpha_c + \alpha_l M_{ps}, \nonumber \\
	f_s(M_{ps}) &=& \beta_c + \beta_s M_{ps}^2, \mbox{~~~and} \nonumber \\
	f_q(M_{ps}) &=& \theta_c + \theta_l M_{ps} + \theta_s M_{ps}^2.
\eeqa{massdep}

\begin{figure}[h]
\includegraphics[scale=0.535]{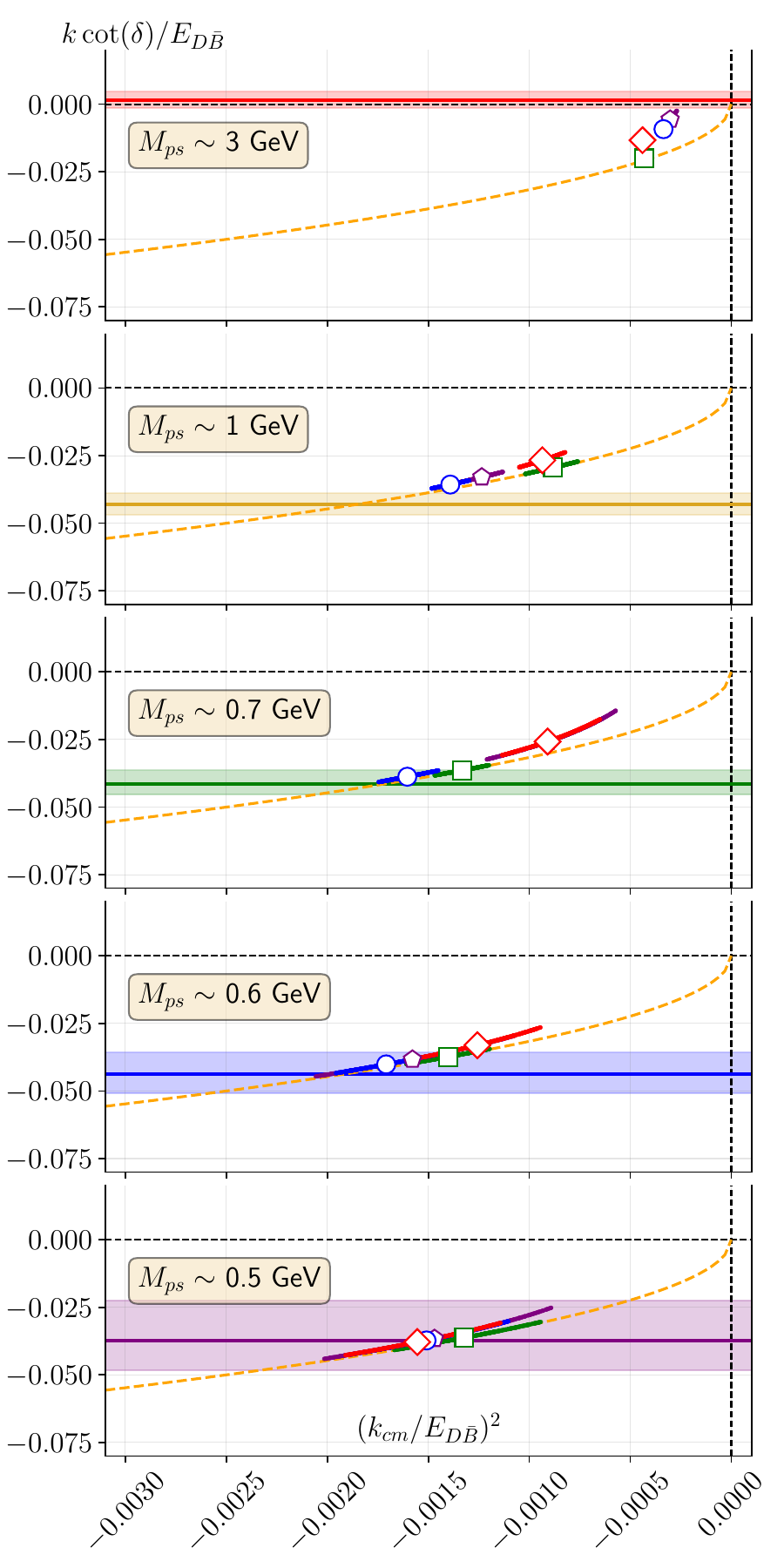}
\caption{$k\cot[\delta_0]$ versus $k^2$ for the different $M_{ps}$ values shown in the legend. 
The scales in either axis are plotted in units of elastic $D\bar{B}$ threshold $E_{D\bar{B}}$. The dashed orange 
curve is the unitarity parabola related to the existence of a real bound state pole 
in the scattering amplitude. The horizontal bands are the continuum extrapolated amplitudes in 
\eqn{pcotd} for each $M_{ps}$, also listed in~\tbn{Ampfits1}.}
\eef{pcotdelta_boot}

\begin{figure}[h]
\includegraphics[scale=0.55]{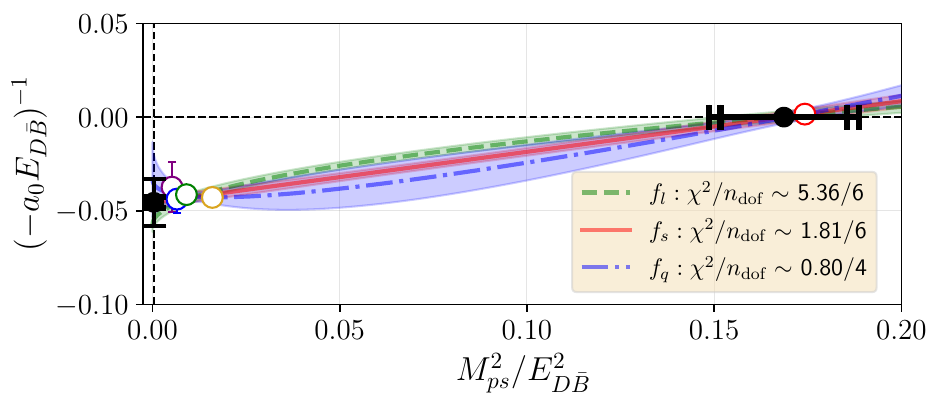}
\caption{The elastic $D\bar{B}$ scattering amplitude in the $S$-wave as a function of the light quark mass, 
in terms of the pseudoscalar mass squared $M_{ps}^2$. The amplitude and $M_{ps}^2$ are presented in 
units of the energy of $D\bar{B}$ threshold ($E_{D\bar{B}}$). The bands indicate fits with different functional 
dependencies listed in \eqn{massdep}. The vertical dotted line near the y-axis represents the 
physical pion mass and the black star on it indicates the scattering amplitude in the physical limit. 
Another star symbol on the $x$-axis indicates the critical $M_{ps}$ where the $D\bar{B}$ system becomes unbound.}
\eef{a0dep}

The light quark mass dependence is determined by minimizing another cost function. The function 
is defined in 
terms of the differences in the data with the phenomenologically motivated parameterizations 
(\textit{c.f} \eqn{massdep}) for its $M_{ps}$ dependence and the data covariance. We present the results 
for this quark mass dependence in \fgn{a0dep} together with the lattice extracted amplitudes as a 
function of $M_{ps}^2/E_{D\bar{B}}^2$. The two black symbols represents amplitude at the physical pion mass limit 
($y$-axis intercept; $M_{ps}=M_{\pi}$) and the critical mass ($x$-axis intercept; $M_{ps}^*$) at which 
the system is close to unitarity branch point. The inner errors associated with these black symbols represent 
the statistical errors, whereas the outer errors also include systematic uncertainties added in quadrature.

The scattering length at the physical pion mass $M_{ps}=M_{\pi}$ 
\beq
a_0^{phys} = 0.61(^{+3}_{-4})(18) \mbox{~fm}
\eeq{scatlen}
together with the observed negative energy shifts in the interacting lattice levels indicate an attractive 
interaction between the $B$ and $D$ mesons, similar to the observation in Ref.~\cite{Padmanath:2023rdu}. 
This attraction is sufficiently strong enough to hold a real bound state with a binding energy
\beq
\delta E_{T_{bc}} = E_{T_{bc}}-E_{D\bar{B}} = -39(^{+4}_{-6})(^{~+8}_{-18}) \mbox{~MeV}.
\eeq{betbc}
When $m_{u/d}>>\Lambda_{QCD}$, the leading linear behavior in $M_{ps}$ is expected to be a good 
description. The black star at the $x$-axis intercept based on the linear $M_{ps}$ dependence 
in \fgn{a0dep} indicates the critical point 
\beq
M^{*}_{ps} = 2.94(15)(5) \mbox{~GeV}.
\eeq{unbound}
at which $a_0$ changes its sign from negative to positive. $M^{*}_{ps}$ and the associated errors are 
evaluated from the fit form $f_l(M_{ps})$ inspired by the leading linear behavior based on heavy quark 
effective field theory \cite{Neubert:1993mb}. Note that the inverse scattering length at the charm 
point is consistent with zero and any fit form is constrained by the data at the charm point. Hence 
systematics associated with the critical mass estimates are significantly small compare to the statistical 
errors.

\bet[tbh]
  \begin{center}
  \begin{tabular}{ l   c }
  \hline \hline
  Source& Error [fm]$\times10^2$ \\\hline
  \multirow{2}{*}{Statistical Errors} & \multirow{2}{*}{\large$\left(^{+3}_{-4}\right)$}\\
   \\\hline
    scale setting  & 3 \\ 
    $m_{b/c}$ tuning & 3 \\
    excited states & 4  \\
    continuum extrapolation & 8 \\
    chiral extrapolation & 15 \\
   \hline
   Total systematics & 18 \\ 
   \hline
  \end{tabular}
  \end{center}
  \caption{The error budget in the calculation of the scattering length, $a_0^{phys}$. This includes the systematics 
  involved as a result of scale setting, excited state effects, heavy quark mass tuning, and uncertainties related 
  to chiral and continuum extrapolations. The total systematics is determined by adding differential estimates in quadrature. }
\eet{errb}

\subsection{Systematic Uncertainties}
In this section we discuss various sources of uncertainties in this calculation that are summarized
in~\tbn{errb}. We follow the bootstrap procedure to carefully carry the statistical errors.
The most dominant systematics are observed to be associated with the light quark mass dependence in 
the chiral regime. Different chiral extrapolation fit forms lead to different estimates for 
the physical scattering length more significant than the statistical precision. The combination of 
$N_f=2+1+1$ MILC lattice QCD ensembles we employ, together with the partially quenched setup using 
an overlap fermion action for light and charm quarks, and an NRQCD formulation for bottom quarks, 
and a rigorous heavy quark mass tuning procedure has been demonstrated to be quite efficient in 
extracting the ground states from finite volume. This setup also reproduces the $1S$ hyperfine 
splittings in quarkonia very precisely with uncertainties less than 6 MeV \cite{Mathur:2022ovu,
Mathur:2016hsm}. The energy splittings and mass ratios we have adopted to work with, efficiently 
mitigate the systematics associated with the lattice realization of heavy quark dynamics \cite{Mathur:2018epb,
Mathur:2022ovu}. We have also included the errors due to fit-window which includes the excited-state 
contamination. The values within the second parenthesis in Equations \ref{scatlen}, \ref{betbc}, and 
\ref{unbound} represent the cumulative systematic uncertainties added in quadrature where the uncertainties 
arising from chiral extrapolation fit forms can be observed to be dominant from \tbn{errb} \cite{Mathur:2018epb,Mathur:2022ovu}.

\section{Discussion on the bindings of $T_{bc}$}
\label{comparison}
\begin{figure*}[htb!]
        \includegraphics[scale=0.6]{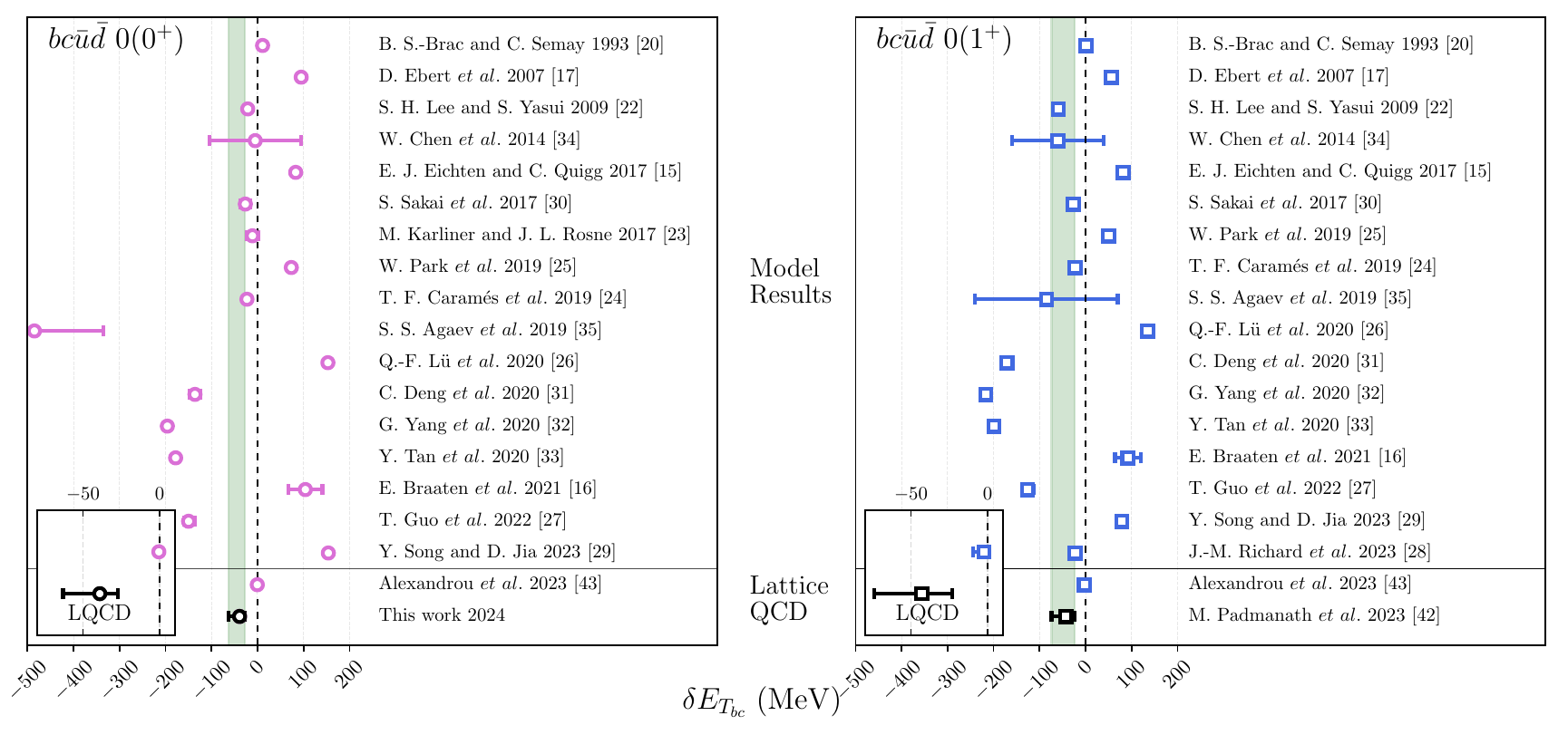}
        \caption{The binding energy calculated in this work in comparison to the recent lattice QCD 
        calculation~\cite{Alexandrou:2023cqg} and other non-lattice determinations. Estimates from 
        non-lattice approaches seem to have a mixed conclusion where several of them show shallow/deep
        binding and many others predicting an unbound state.}
        \label{fg:comp}
\end{figure*}

At this stage, it is natural to assess, where our results stand among other existing lattice QCD-based and 
phenomenological calculations of $D\bar B$($D\bar B^{*}$) scattering in the isoscalar channels. Our investigations 
presented in this work (Ref.~\cite{Padmanath:2023rdu}) indicate negative finite-volume energy shifts 
in S-wave elastic scattering in $D\bar B$($D\bar B^{*}$) meson systems. Further analysis of scattering amplitude 
using finite volume L\"uscher method points to the existence of a real square integrable bound state with 
binding energy of approximately 40 MeV in scalar (see \eqn{betbc}) and axialvector channels in 
Ref.~\cite{Padmanath:2023rdu}. While the erorrbar is large in the estimation in the binding energy, 
the conclusion on the attractiveness is robust. Recently another lattice QCD calculation with a different 
lattice setup has also confirms the attractive nature of interactions in both the channels, however, with 
a much lower binding, just below the respective threshold energies~\cite{Padmanath:2023rdu}.

In Figure \ref{fg:comp}, we present the results from various calculations on the binding of $T_{bc}$ that have 
been predicted over the years. The results presented include those determined using lattice QCD and the non-lattice 
methods, separated by a horizontal line. In each plot, the vertical dashed lines are the respective elastic 
thresholds ($D\bar B$ for $0^+$ and $D\bar B^*$ for $1^+$). Results to the left of this vertical line suggests 
a bound state, whereas those lying to the right points to an unbound system. The vertical green bands are 
the results from our calculations (left: this work, right is from Ref.~\cite{Padmanath:2023rdu}) in perspective 
to those of other calculations. The left plot 
shows the results for $0(0^+)$ channel while the right one is for those of $0(1^+)$ channel. Estimates from 
non-lattice approaches seem to have a large spread of the order of several hundred MeV across the threshold. 
Both the lattice QCD results point towards the existence of bound states of $T_{bc}$. However, more detailed lattice 
calculations are necessary to find the exact locations and the nature of the bound state poles. Given these 
predictions from lattice QCD calculations, and considering the importance of the $T_{bc}$ states as discussed 
in the introduction, experimental searches for these states would indeed be highly worthwhile in the near future.

Another interesting quantity to compare is the scattering length determining the small momentum meson-meson 
interactions in different doubly heavy quark systems ($T_{bb}$, $T_{bc}$ and $T_{cc}$) across various LQCD 
calculations. On the left hand side of Figure \ref{fg:comp_lat}, we present the inverse scattering length ($1/a_0$) 
at the physical $M_{ps}$ determined for these three exotic systems from different lattice calculations \cite{Leskovec:2019ioa,
Lyu:2023xro,Aoki:2023nzp,Padmanath:2023rdu}, where the~\cite{Padmanath:2023rdu} is our previous study using 
the same setup as the present study, together with the scattering length for the discovered $T_{cc}$ \cite{LHCb:2021vvq}. 
The only other LQCD study of $D\bar B$ scattering~\cite{Alexandrou:2023cqg} has also been included (faded point) 
for completeness albeit the analysis not being extrapolated to physical pion mass and to the continuum limit. 
The subscripts $(H)$ and $(L)$ in the $x$-axis tick labels refer to two distinct procedures, the HALQCD and 
L\"uscher-type finite-volume prescription followed respectively, in extracting the scattering length. The HALQCD 
procedure followed in Refs. \cite{Lyu:2023xro,Aoki:2023nzp} provides quite precise estimates, whereas the large 
uncertainty in the $BB^*$ scattering using L\"uscher-type procedure obscures extracting a possible trend, if any 
exist. Subduing these uncertainties require more finite-volume energy levels to constrain the amplitudes, which 
can be achieved either by extracting higher excited states, or by studying more ensembles at different volumes 
or at nonzero lab frame momenta \cite{Alexandrou:2024iwi}. In short, more followup studies involving rigorous 
L\"uscher-type finite-volume treatments with precise estimates are highly desirable to make concrete 
procedure-independent statements on the bindings in different doubly heavy systems. A similar comparison of 
the scattering length in the S-wave scalar $D\bar B$ channel is shown on the right panel of Figure~\ref{fg:comp_lat}.

Considering the differences in systematics between the two evaluations (this work and Ref.~\cite{Alexandrou:2023cqg}) 
for the $D\bar B^{(*)}$ systems, it is too early to argue on the reasons for the observed discrepancies in the magnitude 
of scattering length and binding energy. It could possibly be related to the fact that the results from Ref.~\cite{Alexandrou:2023cqg} lacks any chiral or continuum 
extrapolations or related to the lack of access to the excited elastic excitations in our work, which needs 
to be investigated further. Despite this discrepancy in the magnitude of scattering length, either calculations 
support attractive interactions in these systems. The large errors from our current study naturally indicate equally 
large uncertainty in the binding energy, however the fact that a bound state is seen is expected to be robust and 
consistent given the statistical relevance. Here again, more followup studies with a large number of interpolating 
operators and large statistics with rigorous L\"uscher finite-volume analysis is highly desirable to obtain precise results.

\begin{figure*}[htb!]
        \includegraphics[scale=0.56]{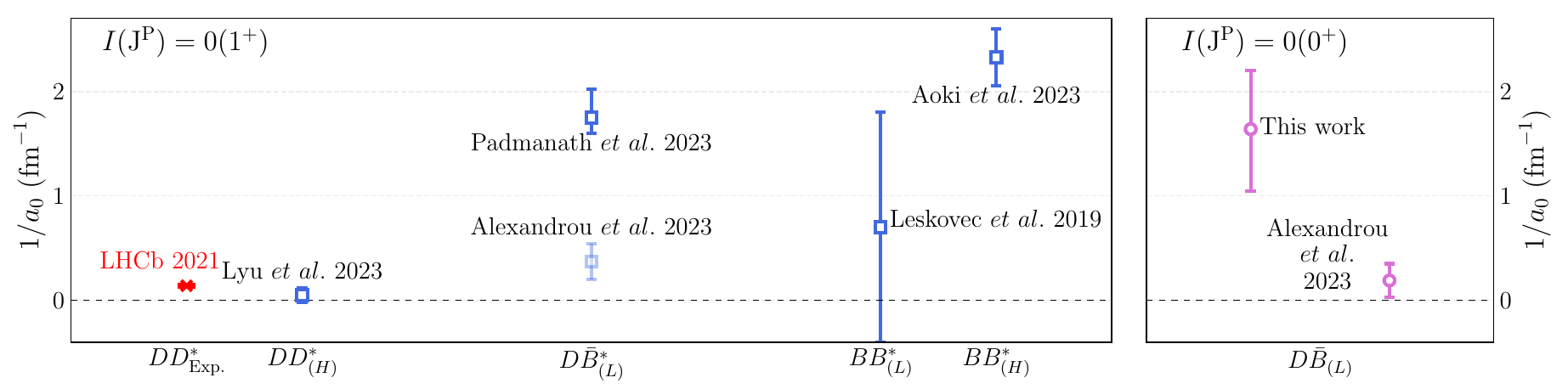}
        \caption{Left plot: The inverse scattering length for $DD^*$, $D\bar B^*$ and $BB^*$ scatterings at
        the physical pion mass as determined in Refs. \cite{Lyu:2023xro,Padmanath:2023rdu,Leskovec:2019ioa,Aoki:2023nzp}.
        The faded point corresponds to a recent lattice evaluation at an unphysically heavy pion mass \cite{Alexandrou:2023cqg}.
        Right plot: The inverse scattering length ($1/a_0$) in $D\bar{B}$ scattering compared between this work and
        Ref.~\cite{Alexandrou:2023cqg}.}
        \label{fg:comp_lat}
\end{figure*}

\section{Summary}
        \label{discuss}
In this work, we present a lattice QCD simulation of elastic S-wave $D\bar{B}$ scattering with explicitly 
exotic flavor $bc\bar u\bar d$ in the isoscalar scalar quantum numbers [$I(J^P) = 0(0^+)$].
We use four $N_f=2+1+1$ ensembles with dynamical Highly Improved Staggered Quark (HISQ) action 
generated by the MILC collaboration, with the valence quarks, up to the charm quark mass, realized 
using an overlap fermion action. The valence bottom quarks are described using an improved NRQCD formulation.

Using the ground state energy levels, presented in Figure~\ref{fg:spec}, we perform a rigorous 
finite-volume amplitude analysis using L\"{u}scher's prescription. The analysis accounts for 
the lattice spacing effects by parameterizing the amplitude with a lattice spacing dependence, 
and taking the continuum limit separately for the five light quark masses studied. The quark mass 
dependence is then investigated to determine the elastic $D\bar{B}$ scattering length $a_0^{phys}$ at 
the physical pion mass and the critical pseudoscalar mass $M^{*}_{ps}$ at which $a_0$ diverges. 
The negative energy shifts in the ground state finite-volume energies taken together with 
the positive estimates for $a_0^{phys}$ (presented in equation~\ref{scatlen}) suggests an attractive 
interaction between the $D$ and $\bar B$ mesons, that is strong enough to form a real square integrable 
bound state with binding energy of $-39(^{+4}_{-6})(^{~+8}_{-18})$ MeV.

Recently another lattice QCD calculation on the $D\bar{B}$ systems also supports an attractive interaction 
between the mesons, however, with a smaller binding and closer to the threshold \cite{Alexandrou:2023cqg}.
Note that this calculation employed bilocal two-meson-type operators at the source and sink and in 
extracting the relevant elastic excitations in the $D\bar{B}$ channel. However, the investigation is limited 
to two lattice ensembles with approximately similar lattice spacings ($\sim 0.12$ fm), that is comparable 
to our coarsest lattice. The apparent discrepancy in the binding energy, whether it is a result of 
uncontrolled excited state contamination due to an asymmetric setup or if it is a result of uncontrolled 
discretization effects, remains to be understood. We leave this issue for the future lattice investigations. 

In this study we are limited to rest frame ground states. While we 
are able to extract the amplitude with a zero-range approximation, future investigations with more rigor in 
extracting elastic excitations are necessary to constrain the energy dependence of the amplitude over a wider
energy range. This would require meson-meson operators with zero overall momentum but individual momentum 
projected mesons like in Ref.~\cite{Alexandrou:2023cqg} to extract the elastic excitations as well as meson-meson 
operators with non-zero overall momentum. Inclusion of such operators is beyond the scope of our current setup. 
Additionally, future studies involving fully dynamical simulations on a wider range of ensembles with different 
fermion actions, high-statistics studies with lighter up and down quark masses, and other improvements. These 
additional efforts would help constrain the relevant scattering amplitude in a framework-independent manner.
In that journey, our calculation is an important step ahead where we have clearly shown the presence of an attractive 
interaction within the $D\bar{B}$ system, with controlled cut-off uncertainties and finite volume effects. Our findings 
offer a stride towards understanding the existence of $0(0^+)$ $T_{bc}$, which could well be the next doubly-heavy 
bound tetraquark to be discovered in the near future.

\begin{acknowledgments}
This work is supported by the Department of Atomic Energy, Government of India, 
under Project Identification Number RTI 4002. M.P. gratefully acknowledges support 
from the Department of Science and Technology, India, SERB Start-up Research Grant No. SRG/2023/001235. 
We are thankful to the MILC collaboration and in particular to S. Gottlieb for providing us with the HISQ lattice ensembles. 
We thank the authors of Ref. \cite{Morningstar:2017spu} for making the {\it TwoHadronsInBox} package utilized in this work. Computations were carried out on the Cray-XC30 of ILGTI, TIFR. N. M. would also like to 
thank A. Salve and K. Ghadiali for computational support. 
\end{acknowledgments}

\appendix


\bibliography{paper}


\end{document}